\documentstyle[twoside,fleqn,espcrc2]{article}
\def\PLB{{\em Phys. Lett. }B }  

\def\PRD{{\em Phys. Rev. }D }

\def\dx{\int d^2x\ \sqrt{-g}\ }
\def\xp{x^+}\def\xm{x^-}
\def\l{\lambda}
\def\ha{{1\over 2}}
\def\a{\alpha}
\def\ha{1\over 2}
\def \ma{{2 M\over \l}}
\def\na{\nabla}

\def\f{\phi}

\def\m{\mu}
\def\n{\nu}
\def\h{\hat}
\def\gmn{g_{\m\n}}
\def\hgmn{\hat \gmn}
\def\dx{\int d^2x\ \sqrt{-g}\ }
      
\def\r{\rho}

\def\t{\tau}
\def\be{\begin{equation}}
\def\ee{\end{equation}}
\def\bea{\begin{eqnarray}}
\def\eea{\end{eqnarray}}
\def\bc{\begin{displaymath}}
\def\ec{\end{displaymath}}
\def\lb{\label}

\newcommand{\AmS}{{\protect\the\textfont2
  A\kern-.1667em\lower.5ex\hbox{M}\kern-.125emS}}

\title{Trace anomaly and Hawking effect in 2D dilaton gravity theories}

\author{Mariano Cadoni\address{Dipartimento di Scienze Fisiche, 
Universit\'a di Cagliari, \\ 
Via Ospedale 72, I-09100 Cagliari, Italy\\
and Istituto Nazionale di Fisica Nucleare, Sezione di Cagliari,
I-09100 Cagliari, Italy}
\thanks{Talk given at the Second Conference on Constrained Dynamics and 
Quantum Gravity, Santa Margherita Ligure, Italy, September 1996, to appear
in the Proceedings}}  

\begin{document}

\begin{abstract}
We investigate the classical and semiclassical features of generic 2D, 
matter-coupled,  
dilaton gravity theories. In particular, we show that the mass,
the temperature and the  flux of Hawking radiation associated with 
2D black holes are invariant under dilaton-dependent Weyl rescalings 
of the metric.  The relationship 
between quantum anomalies and Hawking radiation is discussed.    
\end{abstract}

\maketitle

\section{INTRODUCTION}

These notes summarize the content of two of  our papers  dealing with 
two-dimensional (2D) dilaton gravity theories \cite{CA1,CA2}. 
Our presentation in 
Santa Margherita was essentially based on the first paper \cite{CA1}, 
whereas the content of the second one \cite{CA2} was briefly sketched as 
work in progress. For sake of completeness it seems to us advisable  to 
give a full account  of the recent results of Ref. \cite{CA2} in these notes.

2D dilaton gravity theories have become 
very popular in recent years. They represent simple toy models for 
studying  4D black hole physics and its related challenging issues
such us the ultimate fate of black holes or the loss of quantum 
coherence in the 
evaporation process. Moreover, the relation with non-critical string 
theory and the renormalizability of gravity in two space-time 
dimensions give 
to these theories an intrinsic interest, which goes beyond  black hole 
physics. Even though 2D dilaton gravity contains, as particular cases, models 
with different features (e.g. the Callan-Giddings-Harvey-Strominger  (CGHS) 
model \cite{CGHS}, the Jackiw-Teitelboim (JT) model \cite{JT}), 
a simple, unified description 
of the general theory still exists. 
In this paper we will discuss three crucial  
issues of 2D dilaton gravity, which, as we shall see,  are deeply 
interconnected. These issues are the equivalence of 2D dilaton 
gravity models
under Weyl rescalings of the metric, the existence of black hole 
solutions and the relationship between quantum anomalies and Hawking 
radiation. 

This paper is structured as  follows: In Sect. 2 we  
consider the generic, matter-coupled, classical 2D dilaton gravity  
theory. In particular, we study the behaviour of  physical 
observables under Weyl transformations and the black hole solutions
of the theory. In Sect. 3 we investigate  the semiclassical theory. 
In Sect. 4 we discuss some  particular cases of the generic model.
In Sect. 5 we present our conclusions.   
   
\section{CLASSICAL 2D DILATON GRAVITY}

The most general action of 2D dilaton gravity, conformally coupled to 
a set on $N$ matter scalar fields has the form \cite {MA}    
\bc
S[g,\f,f] ={1\over 2\pi}\dx [D(\f) R(g)
\ec
\be\lb{e1}
+H(\f)(\nabla \f)^{2}
+\l^2 V(\phi)-{\ha} \sum^N_{i=1}(\na f_i)^2 ],
\ee
where $D,H,V$ are arbitrary functions of the dilaton $\f$ and $\l$ is 
a constant. 
An important issue in the context of 2D dilaton gravity is 
the equivalence of models 
connected by conformal  transformations of the metric.
From a purely field theoretical point of view, performing 
dilaton-dependent Weyl rescalings of the metric in the 
2D dilaton gravity action
we should get 
equivalent models, since these transformations are nothing but 
reparametrizations of the field space. The space-time interpretation of 
this equivalence presents, however, some problems because 
geometrical objects such us the scalar curvature of the space-time or 
the equation for the geodesics are not invariant under Weyl 
transformations. 
Let us consider the following Weyl transformation of the metric
\be
\gmn= e^{P(\f)} \hgmn,
\lb{e2}
\ee
where the function $P$ is constrained  by 
requiring  the transformation (\ref{e2}) to be non-singular 
and invertible.
Whereas the matter part of the action (\ref{e1}) is invariant under 
the transformation (\ref{e2}), the gravitational part is not, but it
 maintains its form,  with the functions $D,H,V$ transforming as  ($'=d/d\f$)

\be\lb{e3}
{\h D}=D,\quad
{\h H}=H +D'P',\quad
{\h V}=e^{P}V.
\ee 

The transformation laws (\ref {e2}) and (\ref{e3}) 
enable us to find out how the physical parameters characterizing the 
solutions of the theory transform under the  Weyl transformation (\ref {e2}).
Following Mann \cite {MA}, one can define the conserved quantity 
\bc
M={1\over 2\l} \int^{\f}dD  V\exp\left(-\int d\t{H(\t)\over 
D'(\t)}\right) 
\ec
\be\lb{e4}
- {1\over 2\l} (\na D)^{2} \exp\left(-\int d\t{H(\t)\over 
D'(\t)}\right).
\ee
$M$ is constant whenever the equations of 
motion are satisfied and,  in this case, it can be interpreted as the mass 
of the solution. Using  Eqs. (\ref{e2}), (\ref{e3}), 
 one can easily demonstrate that the mass $M$ given by
the expression (\ref{e4}) is invariant under Weyl transformations of 
the metric.

Choosing $P=-\int^{\f}d\tau \left[H(\tau)/ D'(\tau)\right]$, we can 
always achieve ${\h H}=0$.
The generic static 
solutions in this conformal frame 
have already been found in Ref. \cite {MK}, 
\bc
{\hat{ds}}^{2}=\left(\h J-{2M\over \l}\right)dt^{2}+
\left(\h J-{2M\over \l}\right)^{-1}dr^{2},
\ec 
\be\lb{e5}
\h D(\f)=\l  r,
\ee
where $d\h J/d{\h D}={\h V}$  and $M$ 
is the mass of the solution given by Eq. (\ref{e4}). 
If the equation $\h J=2M/\l$ has at least one solution $\f=\f_{0}$, 
with $\h J$ monotonic, one is lead to interpret the solution as a 
black hole.  However, a rigorous proof of the existence of black 
holes involves a detailed analysis of the global structure of the 
space-time. The dilaton $\f$ gives a coordinate-independent notion of 
location and it can therefore be used to define the asymptotic region, 
the singularities and the event horizon of our 2D space-time. 
Moreover, the natural coupling constant of the theory is $D^{-1/2}$
so that we have a natural division of our space-time in a 
strong-coupling region ($D=0$) and a weak-coupling region ($D=\infty$).
These considerations enable us to identify the weak-coupling region 
$D=\infty$ as the asymptotic region of our space-time.
This notion of 
location is Weyl-invariant because $D$ behaves as a scalar under the 
Weyl transformation (\ref{e2}). 
One can now write down a 
set of conditions that, if satisfied, makes the interpretation of the 
solution 
(\ref{e5}) as a black hole meaningful. One weakness of this kind of 
approach is that we need to use the scalar 
curvature $R$ 
to define the singularities and the asymptotic behaviour of the 
space-time. $R$ is not  Weyl rescaling 
invariant  and cannot be taken as a good quantity for a conformal
invariant characterization of black holes. One has to perform the 
analysis in a particular conformal frame. In  the conformal frame 
defined by Eq. (\ref{e5})  
black holes exist provided the function $\h V$ behaves 
asymptotically as \cite{CA1}
\be\lb{e6} 
\h V\sim D^{\a}, \quad -1<\a\le 1.
\ee
A broader class of models whose static solutions can be interpreted 
as black holes can be obtained choosing a conformal frame in which 
the metric is asymptotically Minkowskian \cite{CA2}.

The Hawking temperature of the generic black hole solution of the 
action (\ref{e1}) is given by
\be\lb{e7}
T={\l \over 4\pi} K(\f_{0}),
\ee
where
\be\lb{e8}
K(\f)=V(\f)\exp\left(-\int^{\f} d\tau {H(\tau) 
\over D'(\tau)}\right).
\ee
The temperature is invariant  under Weyl transformations.
This can be easily checked 
using Eqs. (\ref{e3}) in Eq. (\ref{e8}) and taking 
into account that the transformations (\ref{e2}) do not change 
the position $\f_{0}$ of the event horizon. It is interesting to note 
that the mass (\ref{e4}) and the temperature  (\ref{e7}) are 
invariant not only under Weyl transformations but also under 
reparametrizations of the dilaton field.

\section{QUANTUM ANOMALIES AND HAWKING RADIATION}
It is well-known that in 
quantizing the scalar matter fields $f$ in a fixed background geometry 
 the Weyl rescaling  and/or part of the diffeomorphism  invariance of the 
classical action for the matter fields has to be 
explicitly broken. If one 
decides to preserve diffeomorphism invariance the semiclassical action 
is given by the usual Liouville-Polyakov action but one has still the freedom 
of adding local, covariant, dilaton-dependent counterterms to the 
semiclassical action \cite {ST,RST}. In the path integral formulation
this ambiguity  is related to the choice of the metric to be used in the 
measure. The semiclassical action has the general form
\bea\lb{e9}
S=S_{cl}-{N\over 96\pi}\dx[ R{1\over \nabla^2} R\nonumber\\
-4 N(\phi) R +4G(\phi) (\nabla\phi)^2].
\eea
 The second term above is the usual non-local
Liouville-Polyakov action, $S_{cl}$ is the classical action (\ref{e1}) and 
$N(\phi),G(\phi)$ are two arbitrary functions. The form of the 
functions $N$and $G$ depends on the conformal frame we choose, but, as 
we  shall see later on in this paper, the final result for the Hawking 
radiation rate is invariant under Weyl transformations. In the 
conformal frame where the metric is asymptotically Minkowskian, $N$ 
and $G$ can be fixed requiring 
the expectation value of the stress-energy tensor to vanish when 
evaluated for the $M=0$ ground state solution  (Minkowsky   
space) \cite{CA2}. 

The black hole radiation can now  be studied working in the 
conformal gauge $ds^{2}=-e^{2\rho}dx^{+}dx^{-}$ and 
considering a black hole 
formed by collapse of a $f$-shock-wave, travelling in the $x^{+}$ 
direction and described by a classical stress-energy tensor 
$T_{++}=M\delta(x^{+}-x^{+}_{0})$.
The classical solution describing the collapse of the 
shock-wave, for $\xp\le\xp _0$, is given by
\bc
e^{2\r}=K\exp\left(-\int^{\f} d\tau {H(\tau)\over 
D'(\tau)}\right),
\ec

\be\lb{f1}
\int^\f{d\t\over K(\t)}={\l\over 2}(\xp-\xm),
\ee
and, for $\xp\ge\xp_0$, it is given by 
\bc
e^{2\r}=\exp\left(-\int^{\f} d\tau {H(\tau)\over 
D'(\tau)}\right) \biggl(K-\ma\biggr)F'(\xm),
\ec
\bc
\int^\f{d\t\over K(\t)-\ma}={\l\over 2}
\left[\xp-\xp_0 -F(\xm)\right],
\ec
\be\lb{f2}
F'(\xm)={dF\over d\xm}=\biggl({K\over 
{K-\ma}}\biggr)_{\xp=\xp_0},
\ee
 where $K$ is given as in Eq. (\ref{e8}). The next step in our semiclassical  
calculation is to use the 
effective action (\ref{e9}) to derive the expression for the quantum 
contributions of the matter to the stress-energy tensor. The flux of 
Hawking radiation across spatial infinity is given by $<T_{--}>$ 
evaluated on 
the asymptotic $D=\infty$ region.
For the class of models in Eq. (\ref{e6}) and for the shock-wave 
solution described previously a 
straightforward calculation leads to  \cite {CA2}
\be\lb{f3}
 < T_{--}>_{as}={N\over 24}{1\over (F')^2}\{F,\xm\},
\ee
where $\{F,\xm\}$ denotes the Schwarzian derivative of the function 
$F(\xm)$.
This is a Weyl rescaling and dilaton reparametrization invariant 
result for the Hawking flux. 
In fact the function $F(\xm)$ is defined entirely in terms of the 
function $K(\f)$ (see Eq. (\ref{f2})), which in turn is invariant 
under both 
transformations (see Eq. (\ref{e8})).
Though the trace anomaly is Weyl rescaling 
dependent, the Hawking radiation seen by an asymptotic observer is 
independent of the particular conformal frame chosen.      
When the horizon $\f_{0}$ is approached, the Hawking flux reaches the thermal 
value 
\be\lb {f4}
< T_{--}>_{as}^{h}={N\over 12}{\l^{2}\over 16}[K(\f_{0})]^{2},
\ee
which is the   result found in Ref. \cite {CA1}, written 
in a manifest Weyl rescaling and dilaton 
reparametrization invariant form.

\section{PARTICULAR CASES}

The general model (\ref{e1})   contains, as particular cases, models that 
have already 
been investigated in the literature. In this section we will show how 
previous results
for the CGHS model and the JT theory can be obtained as particular cases of 
our formulation. 
\subsection{String inspired dilaton gravity}
This is the most popular 2D dilaton gravity model. In its original 
derivation \cite{CGHS} the action has the form (\ref{e1}) with 
\be
D=V=e^{-2\f},\quad H=4e^{-2\f}.
\ee
The model admits asymptotically flat black hole  solutions.
Using Eq. (\ref{e7}), (\ref{f4})   we find for the 
temperature and magnitude of the Hawking effect
\be
T={\l\over 4\pi}, \quad  < T_{--}>_{as}^h={N\over 12}{\l^2\over 16}.
\ee
This result coincides, after the redefinition $\l\to 2\l$ with the CGHS 
result \cite{CGHS}.
 All the 2D dilaton gravity models obtained 
from the CGHS model using Weyl transformation are characterized 
by the  same values of mass, temperature and Hawking radiation rate.
In particular, this is true  for the model investigated in Ref.
\cite{CM}. This  model is characterized by $H=0$ and its 
 black hole solutions are described by a Rindler space-time.
Also the models of Ref. \cite {FR} can be obtained from the CGHS model
through a Weyl transformation of the metric \cite{CA2}.
The black hole solutions of these models have, therefore,
the same values of mass, temperature and Hawking radiation
rate as the CGHS black holes.
      
\subsection{ The Jackiw-Teitelboim theory}

The JT theory is obtained from the action (\ref{e1}) by taking 
$H=0$ and $2D=V=2\exp(-2\f)$.
The model admits black hole solutions  with asymptotic
 anti-de Sitter behaviour. More precisely, as shown in Ref. \cite{CM1},
the black hole space-time is obtained from a particular 
parametrization of 2D anti-de Sitter space-time endowed with a
boundary. 
Eqs. (\ref{e7})  and (\ref{f4})  give now
\bc
T={1\over 2\pi}\sqrt{2M\l}, \quad  < T_{--}>_{as}^h=
{N\over 24}M\l.
\ec
The same result for the Hawking radiation rate has been obtained
in Ref. \cite{CM1}  performing the canonical quantization of the scalar
 fields $f$ in the
anti- de Sitter  background  geometry.

\section {CONCLUSIONS}
In this paper we have discussed  classical and semiclassical 
features of generic 2D dilaton gravity theories. 
We have shown that the usual relationship between conformal
anomalies and Hawking radiation can be extended to a 
broad class of 2D dilaton gravity models. In particular, 
we have found a simple, general formula for
the magnitude of the Hawking effect.
We have also shown that physical observables associated with 
2D black holes such as the mass, the temperature and the hawking
radiation rate are invariant both under Weyl transformations
and dilaton reparametrizations. This implies  
a conformal equivalence between 2D dilaton gravity models.
An important issue we have not discussed in this paper is the 
physical meaning of  this  equivalence. 
 For a detailed 
discussion of this issue see  Ref. \cite {CA2}. 



\begin{thebibliography}{9}
\bibitem{CA1}
M. Cadoni, \PRD  {\bf 53} (1996)  4413.
\bibitem{CA2} 
M. Cadoni, hep-th/9610201.
\bibitem {CGHS}
C.G. Callan, S.B. Giddings, J.A. Harvey and A. Strominger, 
\PRD {\bf 45} (1992) 1005.
\bibitem {JT}
 C. Teitelboim, in {\sl Quantum Theory of gravity}, 
S.M. Christensen,
ed. (Adam Hilger, Bristol, 1984); R. Jackiw, {\sl ibidem}.
\bibitem {MA}
 R. B. Mann, \PRD {\bf 47} (1993) 4438.
\bibitem {MK}
D. Louis-Martinez and G. Kunstatter, \PRD {\bf 49}  (1994) 5227.

\bibitem{ST}
 A. Strominger, \PRD {\bf 46}, 4396 (1992).

\bibitem {RST}
 J.C. Russo, L. Susskind, L. Thorlacius, \PRD {\bf 46} (1992)
3444.

\bibitem {CM}
 M. Cadoni, S. Mignemi \PLB  {\bf 358} (1995) 217.

\bibitem {FR}
A. Fabbri, J.G. Russo, hep-th/9510109.

\bibitem {CM1}
M. Cadoni and S. Mignemi, 
\PRD  {\bf 51} (1995) 4319.
\end{thebibliography}
\end{document}